\documentclass[fleqn,usenatbib]{mnras}
\usepackage{newtxtext,newtxmath}
\usepackage{xspace}
\usepackage{longtable}
\usepackage{subcaption} %
\usepackage{chngcntr}

\usepackage[T1]{fontenc}

\DeclareRobustCommand{\VAN}[3]{#2}
\let\VANthebibliography\thebibliography
\def\thebibliography{\DeclareRobustCommand{\VAN}[3]{##3}\VANthebibliography}

\usepackage{graphicx}   % Including figure files
\usepackage{subcaption}
\usepackage{amsmath}
\usepackage{orcidlink}

\title[A Statistical Analysis on Non-repeating FRBs]
{A Statistical Analysis of Fluence and Energy Distributions of Non-repeating Fast Radio Bursts Detected by CHIME}

\author[Nurimangul Nurmamat et al.]
{Nurimangul Nurmamat \orcidlink{0000-0001-9227-3716} $^{1}$,
 Yong-Feng Huang \orcidlink{0000-0001-7199-2906} $^{1,2}$\thanks{E-mail: hyf@nju.edu.cn},
 Xiao-Fei Dong \orcidlink{0009-0000-0467-0050} $^{1}$,
 Chen-Ran Hu \orcidlink{0000-0002-5238-8997} $^{1}$,
 Orkash Amat \orcidlink{0000-0003-3230-7587} $^{1}$,
\newauthor
 Ze-Cheng Zou \orcidlink{0000-0002-6189-8307} $^{1}$,
 Abdusattar Kurban \orcidlink{0000-0002-2162-0378} $^{3,4}$,
 Jin-Jun Geng \orcidlink{0000-0001-9648-7295} $^{5}$,
 Chen Deng \orcidlink{0000-0002-2191-7286} $^{1}$
\\\\
$^{1}$School of Astronomy and Space Science, Nanjing University, Nanjing 210023, China \\
$^{2}$Key Laboratory of Modern Astronomy and Astrophysics
 (Nanjing University), Ministry of Education, China \\
$^{3}$Xinjiang Astronomical Observatory, Chinese Academy of Sciences, Urumqi 830011, China \\
$^{4}$Xinjiang Key Laboratory of Radio Astrophysics, Urumqi 830011, China \\
$^{5}$Purple Mountain Observatory, Chinese Academy of Sciences, Nanjing 210023, China \\
}

\date{Accepted XXX. Received YYY; in original form ZZZ}

\pubyear{2025}

\begin{document}
\label{firstpage}
\pagerange{\pageref{firstpage}--\pageref{lastpage}}
\maketitle

% Abstract of the paper
\begin{abstract}
Fast Radio Bursts (FRBs) are energetic radio bursts that typically
last for milliseconds. They are mostly of extragalactic origin,
but the progenitors, trigger mechanisms and radiation processes
are still largely unknown. Here we present a comprehensive
analysis on 415 non-repeating FRBs detected by CHIME, applying
manual filtering to ensure sample completeness. It is found that
the distribution of fluence can be approximated by a three-segment
power-law function, with the power-law indices being $-3.76 \pm
1.61$, $0.20 \pm 0.68$ and $2.06 \pm 0.90$ in the low, middle, and
high fluence segments, respectively.  Both the total dispersion
measure (\text{DM}) and the extragalactic \text{DM} follow a
smoothly broken power-law distribution, with characteristic break
DM values of $\sim 703$ pc $\rm cm^{-3}$ and $\sim 639$ pc $\rm
cm^{-3}$, respectively. The redshifts are estimated from the
extragalactic \text{DM} by using the Macquart relation, which are
found to peak at $ z \sim 0.6$. The isotropic energy release
($E_{\text{iso}}$) is also derived for each burst. Two-Gaussian
components are revealed in the distribution of $E_{\text{iso}}$,
with the major population narrowly clustered at $\sim 2.3 \times
10^{40} {\rm erg}$. The minor population have a characteristic
energy of $\sim 1.6 \times 10^{39}$ erg and span approximately one
order of magnitude. The distribution hints a near-uniform energy
release mechanism for the dominant population
as expected from some catastrophic channels,
whereas the lower-energy component (potentially including
repeat-capable sources) may reflect a broader diversity in FRB
origins, emission mechanisms and evolutionary stages. 
\end{abstract}

\begin{keywords}
fast radio bursts -- radio continuum: transients
\end{keywords}

%%%%%%%%%%%%%%%%%%%%%%%%%%%%%%%%%%%%%%%%%%%%%%%%%%

%%%%%%%%%%%%%%%%% BODY OF PAPER %%%%%%%%%%%%%%%%%%

\section{Introduction}
\label{sect:Intro}

Fast Radio Bursts (FRBs), characterized by their extremely short
durations (on millisecond timescales) and intense radio emissions,
were first reported by~\cite{Lorimer2007Sci}. Since the discovery,
over 800 FRBs have been identified and documented to date
~\citep{2021CHIME/FRB, 2022Petroff,2023Xu,2024ApJCHIME/FRB}. Some
of these FRB sources exhibit stochastic repetition patterns
~\citep{Spitler2016,Oppermann2018,2020Chime, CHIME2023}, while
others display quasi-periodic behavior across a broad range of
timescales ~\citep{2022Lanman,PastorMarazuela2023}. Non-repeating
FRBs are not observed to produce bursts for more than one time,
though some share common morphological features with repeating
FRBs in various aspects such as frequency drift, Faraday rotation
measure (RM), and burst frequency
\citep{Michilli2018,Hessels2019ApJ, 2022AJBhandari}. Repeating
FRBs are likely linked to young magnetars, often accompanied by a
persistent radio source, as seen in FRBs 20121102A, 20190520B, and
20201124A ~\citep{Natur2017Chatterjee,
2022NaturNiu,Rahaman2025ApJ}. In fact, statistics from the
CHIME/FRB Collaboration indicates that only $\thicksim 4\%$ of the
CHIME FRBs are confirmed as repeating
sources~\citep{2021CHIME/FRB}. The extragalactic origin of these
non-repeating (or one-off) FRBs was initially inferred from their
large dispersion measures (\text{DM}s) and has since been
conclusively confirmed through direct redshift measurements in
several cases
~\citep{Petroff2019AARv,Cordes2019ARAA,Ravi2019,Bannister2019}.

The trigger mechanism of FRBs still remains unclear. The host
galaxies and local environments of repeating and non-repeating
FRBs differ in some aspects, which implies that their origins and
underlying physical mechanisms may also be somewhat different.
Repeating FRBs are likely linked to young magnetars. They are
predominantly found in low-metallicity, high star-forming dwarf
galaxies, and are more frequently accompanied by a persistent
radio source, as seen in FRBs 20121102A and 20190520B
~\citep{Natur2017Chatterjee,2017ApJMarcote,2022NaturNiu}.
Additionally, repeating FRBs typically have a high Faraday RM and
display a pronounced frequency drift in their dynamical spectra,
indicating a strong local magnetic
field~\citep{Michilli2018,Hessels2019ApJ}. In contrast,
non-repeating FRBs are characterized by a single, short-duration
burst with lower RM values, reflecting a weaker local magnetic
field ~\citep{Ravi2019,Bannister2019,Pandhi2024}. They may result
from catastrophic events, such as neutron star mergers, supernova
explosions, or black hole accretion episodes~\citep{Katz2024}.
They are often found in a more diverse range of host galaxy types,
reflecting the stochastic and heterogeneous nature of their
progenitors ~\citep{Totani2013,Piro2016,Margalit2019}.
Interestingly, \citet{Margalit2019} proposed a unified picture
that both repeating and non-repeating FRBs originate from
magnetars which are at different evolutionary stages: young,
highly magnetized neutron stars produce repeating FRBs, while
older magnetars with weaker magnetic fields are involved in
non-repeating FRBs.  Recent work by \cite{Kirsten2024NatA} also
supports this idea, who found that the energy distributions of
repeating and non-repeating FRBs are similar, suggesting that both
types may indeed share a common origin. Note that various other
FRB models involving compact stars have also been suggested and
could not be expelled yet
~\citep{Platts2019,2020Bochenek,2021Voisin,2021Geng,Kurban,2024EPJC}.
For a comprehensive review, see the FRB Theory
Catalog~\footnote{\url{http://frbtheorycat.org}}.

Observational data play a crucial role in understanding the
properties and origins of
FRBs~\citep{Locatelli2019,Hu2023,Wu2024,Ng2024}. Some key
parameters, such as fluence ($F$), $\mathrm{DM}$, extragalactic DM
($\mathrm{DM}_{\rm exc}$), redshift ($z$), and isotropic energy
($E_{\text{iso}}$), are essential for revealing the intrinsic
characteristics of FRBs~\citep{2023Zhang}. Recent analyses of FRB
fluence distributions have revealed diverse energy release
patterns. For example, ~\cite{James2019} identified a steepening
of the fluence distribution at certain flux threshold, which may
reflect the intrinsic variation in the FRB population if it were
not due to observational bias. Similarly, ~\cite{Zhang2022}
demonstrated that the fluence distribution of highly active
repeating FRBs, such as FRB 20201124A, can be well fitted by a
broken power-law function, highlighting a different energy
pattern.

Similarly, statistical analyses on $\mathrm{DM}$ and $\mathrm{DM}_{\rm exc}$ offer
critical clues about the distances and local environments of FRBs.
By combining $\mathrm{DM}$ with redshift information, researchers can
estimate the contribution of the intergalactic medium (\text{IGM})
and probe the large-scale structures of the universe.
\cite{Zhang2021} studied the isotropic-equivalent energy
distribution of FRBs. They found that it follows a power-law
function with an index of approximately -1.8, which is roughly
consistent with previous
studies~\citep{Luo2018,Lu2019L,Luo2020,Lu2020}.

Non-repeating FRBs provide a unique opportunity to probe one-off
astrophysical events through population-wide statistical studies.
By focusing solely on non-repeating FRBs, we can avoid biases
introduced by the large number of repeated bursts from a single
source, thereby establishing a clearer picture of the
population-wide distribution of FRB parameters. These studies
yield critical insights into FRB energetics, distances, and local
environments, helping constrain potential progenitor models.

Although recent years have seen extensive efforts to model
intrinsic FRB properties, such as redshift evolution, repetition
statistics, and luminosity functions, using modern statistical
inference and population synthesis techniques
~\citep{Connor2019,2022James,Shin2023ApJ,Peng2025,Ma2025}, the
vast majority of bursts still lack directly measured redshifts
($z$) and intrinsic energies ($E_{\mathrm{iso}}$). Under this
observational limitation, large sample studies must continue to
rely on measurable quantities such as $\mathrm{DM}$ and fluence.
Systematically analysing the distributions of these observables is
essential for assessing instrumental and environmental
selection effects and for providing the statistical data
needed to infer intrinsic FRB physics
\citep{Macquart2018,James2019,Cordes2019ARAA}. An earlier study by
~\citet{2017RAALi} presented a similar $\mathrm{DM}$ and fluence analysis,
but its conclusions were limited by a much smaller sample. It is
necessary to leverage a substantially larger data set to deliver
updated distributions of \text{DM}, fluence, redshifts and
energies, furnishing useful statistical inputs and constraints for
future population and cosmological modelling of FRBs.

In this study, we perform a comprehensive statistical analysis on
the observational data of non-repeating FRBs. The distributions of
various observed and derived parameters such as $F$, $\mathrm{DM}$,
$\mathrm{DM}_{\rm exc}$, $z$, and $E_{\text{iso}}$ are analyzed. The
structure of our paper is organized as follows. In
Section~\ref{sect:Obs}, we describe the data sources of our
non-repeating FRB sample. To address potential
selection effects in the CHIME data, we apply filtering criteria
to enhance sample completeness. Section~\ref{sec:3} presents our
results on the fluence distribution. The results on other FRB
parameters, such as distance, redshift and energetics are
presented in Section~\ref{sec:method}. In
Section~\ref{sect:Comparison}, we compare our results with some previous studies.
Finally, Section~\ref{sec:conclusions} presents our conclusions
and discussion.

\section{Data Source and Sample Selection}
\label{sect:Obs}

In this study, we analyze the FRBs observed by the Canadian
Hydrogen Intensity Mapping Experiment
(CHIME)~\footnote{\url{https://www.chime-frb.ca/catalog}}. The
data are accessed via the Blinkverse
platform~\footnote{\url{https://blinkverse.zero2x.org/\#/availability}}
on January 1, 2025. CHIME is a Canadian radio telescope array
designed to map the distribution of hydrogen in the universe. It
can effectively detect FRBs due to its large field of view. The
CHIME Catalog is a useful database containing FRBs observed by the
array. The data in the catalog spans several years, beginning in
2018 July 25. It includes both repeating and non-repeating FRBs.
The \text{DM} ranges from 103 pc cm$^{-3}$ to 3038 pc cm$^{-3}$,
providing insights into the FRB distances and the characteristics
of the intervening medium. The fluence ranges from a few mJy ms to
a hundred Jy ms, and the duration spans from tens of milliseconds
to a few seconds \citep{2021CHIME/FRB}. The signal-to-noise ratio
(\text{SNR}) of most bursts exceeds 10, ensuring a high-quality record of
the events. Observations were made over a frequency range of 400
MHz to 800 MHz, enabling the detection of a diverse set of FRBs.
These data allow us to investigate the physical properties of
FRBs, their environments, and their cosmological distribution.

In our study, we focus on 461 non-repeating FRBs from the
CHIME/FRB Catalog 1~\citep{2021CHIME/FRB}. We chose those events
with necessary parameters available, including the fluence ($F$),
pulse width, $\mathrm{DM}$, Galactic longitude (GL), Galactic
latitude (GB), and central frequency ($\rm \nu_{\rm c}$). Events
with significant gaps or missing data were excluded to maintain
consistency and reliability in our analysis.

To mitigate selection effects such as
telescope sensitivity limits and beam variations that bias towards
brighter bursts ~\citep{2021CHIME/FRB,Hashimoto2022MNRAS}, we
apply a set of manual filtering criteria to the initial 461
events:
\begin{itemize}
  \item $\mathrm{SNR} \geq$ 12, to ensure a reliable detection above the noise threshold.
  \item $\mathrm{DM}_{\rm SNR}$ $\geq$ 1.5 $\times$ max($\mathrm{DM}_{\rm NE2001}$),
   to ensure an extragalactic origin by excluding potential Galactic contributions.
  \item Fluence $\geq$ 0.4 Jy ms, to exclude invalid (zero) or unreliable
  low-fluence events.
\end{itemize}
This yields a more statistically complete subsample of 415 events.
While injection-based methods
~\citep{2021CHIME/FRB,Shin2023ApJ,Cui2025ApJ} could provide
precise corrections for detection efficiencies, they require
access to raw telescope data, proprietary CHIME/FRB simulation
pipelines, and significant computational resources, which are
beyond the scope of this analysis.

Note that here we use the \texttt{fitburst} SNR (the
\texttt{snr\_fitb} column in CHIME/FRB Catalog~1) rather than the
real-time \texttt{bonsai} SNR adopted in some previous analyses
~\citep{Hashimoto2022MNRAS}. The \texttt{bonsai} SNR is the
real-time trigger statistic that might be affected by
radio-frequency interference (RFI) clipping for bright bursts and
does not always represent the true signal amplitude (see
Section~5.3 of \citealt{2021CHIME/FRB}). In contrast, the
\texttt{fitburst} SNR provides a more physically consistent and
morphology-independent measure of burst significance, both for
wide and narrow-band bursts. It performs a full maximum-likelihood
fit to the dynamic spectrum, modeling the dispersion, scattering,
and spectral structure at higher time--frequency resolution (see
Section~3.3 of \citealt{2021CHIME/FRB}). Therefore, using the
\texttt{fitburst} SNR ensures a self-consistent sample selection.

\section{Fluence distribution of non-repeating FRBs}\label{sec:3}

\begin{figure*}
\centering
\includegraphics[width=0.48\textwidth]{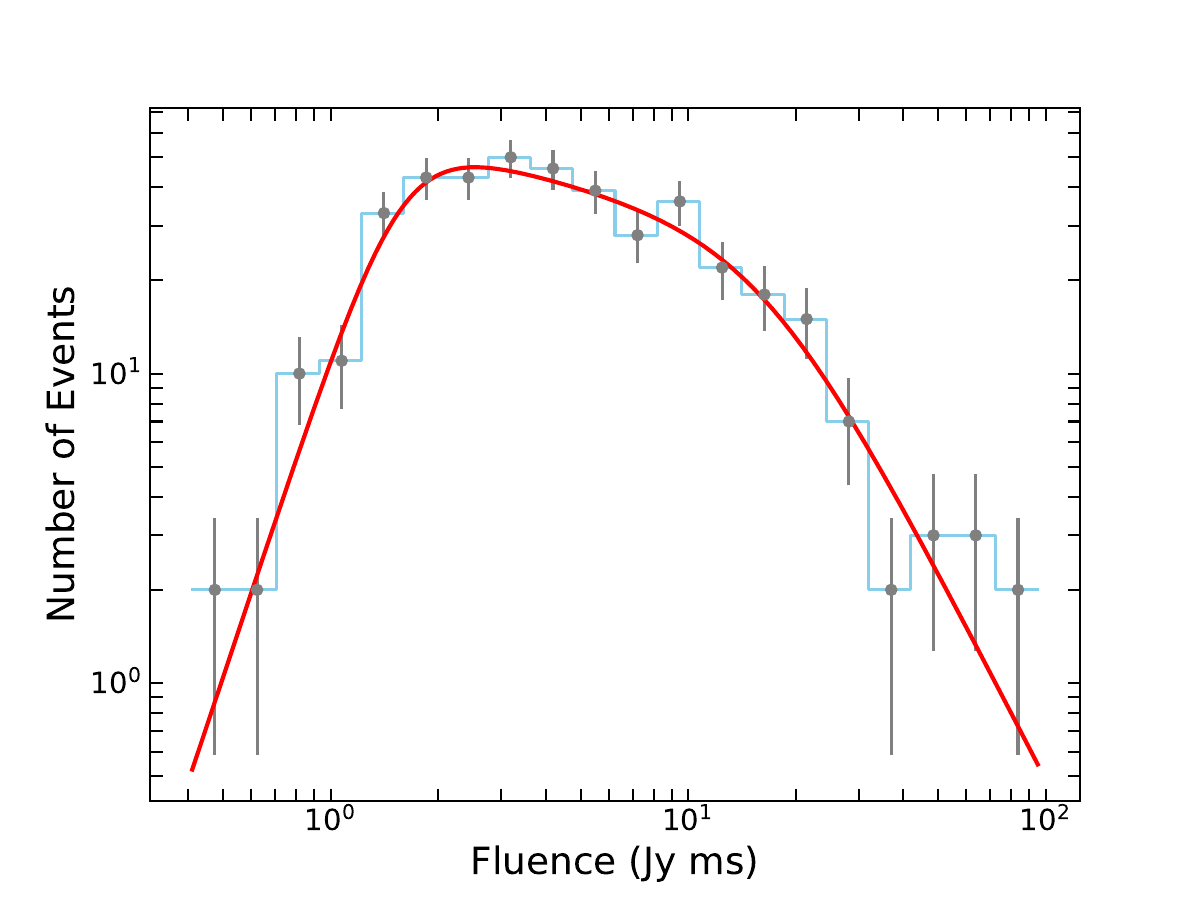}
\includegraphics[width=0.42\textwidth]{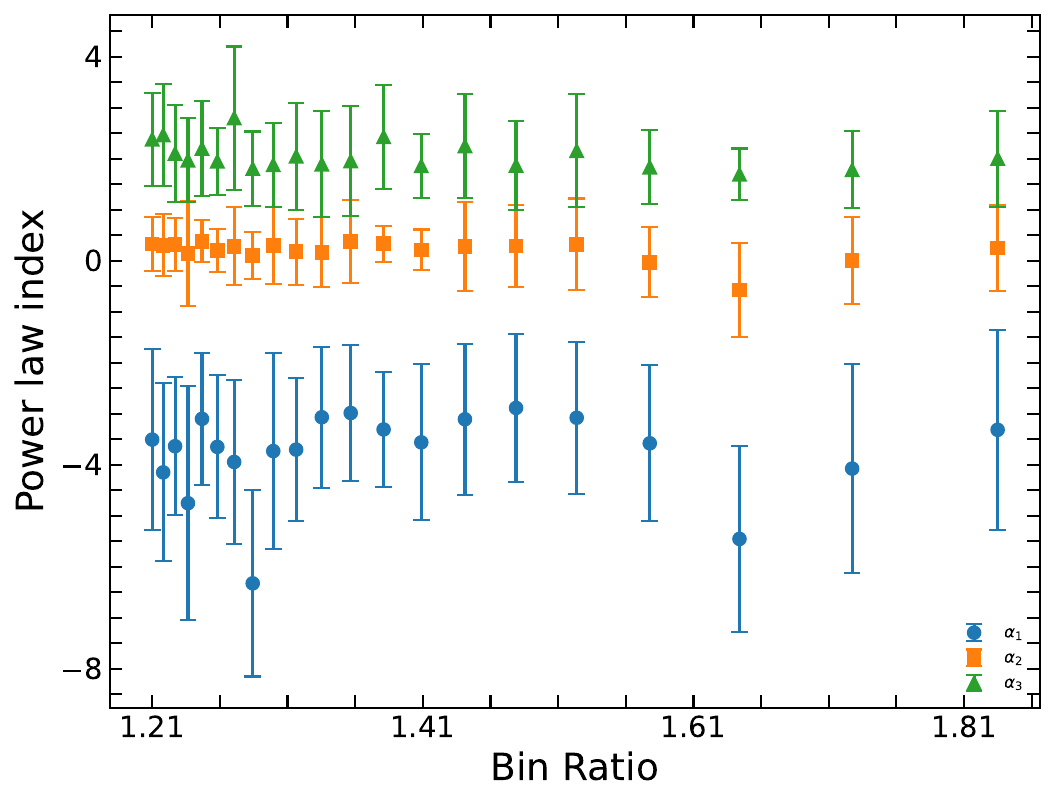}
\caption{The fluence distribution of non-repeating FRBs
(filtered data with $\rm N = 415$ events).
  The left panel shows the the histogram of the observational
  data, with a bin ratio of 1.3. The best MCMC fitting result
  engaging a three-segment power-law function is also illustrated.
  The right panel illustrates the variation of $\alpha_1$,
  $\alpha_2$, and $\alpha_3$ with respect to the bin ratio.}
\label{fig:best-fitted}
\end{figure*}

The exact redshift and distance are unavailable for most
non-repeating FRBs, which means the intrinsic luminosity and
energetics are also unknown. The fluence can be regarded as an
apparent intensity, which tells us how ``bright'' the burst seems
to be. The distribution of fluence is a useful tool for
investigating the physical properties and spatial distribution of
the FRBs, which can help us determine their event rates to some
extent. Additionally, the fluence distribution can also provide
insights into the underlying physical mechanisms driving FRBs,
helping assess whether they can potentially act as standard
candles.

The fluence distribution of the filtered 415
non-repeating FRBs is shown in \autoref{fig:best-fitted} on the
logarithmic scale. The raw histogram visually exhibits three
segments, i.e. a steep rise at low-fluence regime, a slow decline
at middle-fluence regime, and a steep decline at high-fluence
regime. This motivates us to use a three-segment broken power-law
function to fit the histogram, which is expressed as,
\begin{equation}
    N(F) =
    a \left(
    \left( \frac{F}{F_1} \right)^{\alpha_1 \omega} +
    \left( \frac{F}{F_1} \right)^{\alpha_2 \omega} +
    \left( \frac{F}{F_2} \right)^{\alpha_3 \omega}
    \right)^{-\frac{1}{\omega}},
\end{equation}
where $F$ is the fluence, $a$ is a normalization constant, $F_{1}$
and $F_{2}$ are two characteristic fluence values that correspond
to the transition points of the distribution. $\alpha_1$,
$\alpha_2$ and $\alpha_3$ are the power-law indices of the three
fluence segments. $\omega$ is a parameter that features the
smoothness of the transition between two adjacent segments. A
larger $\omega$ indicates a sharper transition, while a smaller
$\omega$ leads to a smoother transition. This function is a
generalization of the simple broken power-law function, enabling a
satisfactory description of a multiple-component distribution with
distinct behaviors in low and high fluence regions.

The Markov Chain Monte Carlo (MCMC) method~\citep{2013asc} is
engaged to get a best fit to the observed fluence distribution by
using Equation (1). In the left panel of
\autoref{fig:best-fitted}, the best-fit curve is also illustrated
for a bin ratio of $\sim 1.3$ for the filtered sample. In this
case, the best-fit parameters are derived as $a = 57.1 \pm 15.6$
(in arbitrary unit), $F_{\mathrm 1} = 1.6 \pm 0.4$ (\rm Jy ms),
$F_{\mathrm 2} = 12.1 \pm 3.7$ (\rm Jy ms), $\alpha_1 = -3.48 \pm
1.25$, $\alpha_2 = 0.30 \pm 0.50$, $\alpha_3 = 2.26 \pm 1.07$, and
$\omega = 1.40 \pm 1.15$. Here the uncertainties are given in
1$\sigma$ (68\%) posterior range, i.e. the standard deviation of
the marginalized MCMC samples (estimated by using \texttt{emcee}
with 32 walkers and 2000 steps, discarding the first 100 as
burn-in).

Due to the limited sample size, we noticed that the best-fit
power-law indices ($\alpha_1$, $\alpha_2$ and $\alpha_3$) are
affected by the bin width. Various bin ratios (which determine the
bin width) were thus adopted and compared in our modeling.
The bin ratio, defined as the ratio of
adjacent bin boundaries in log-space, varies from
1.24 to 1.83 to test the robustness of the fit against binning choices.

In the right panel of \autoref{fig:best-fitted}, the results of
$\alpha_1$, $\alpha_2$ and $\alpha_3$ are plotted when the bin
ratio varies in 1.24 -- 1.83. Generally, it
can be seen that the three indices are relatively stable when the
bin ratio varies in such a wide range. The
average values of the best-fit parameters, derived from MCMC
fitting across multiple bin ratios (1.24 -- 1.83) using filtered
data, are $\alpha_1 = -3.76 \pm 1.61$, $\alpha_2 = 0.20 \pm 0.68$,
and $\alpha_3 = 2.06 \pm 0.90$. The derived parameters are
presented in Table~\ref{tab:best_fit_params}. We notice that the fit to the full
unfiltered sample (see Appendix Table~\ref{tab:full_sample})
yields parameters consistent with those derived from the filtered
sample. They are consistent with each other within 1$\sigma$
range.

The apparent intensity distribution has been studied by several
groups previously~\citep{2017RAALi,Macquart2018}. For example,
using a sample of 16 non-repeating FRBs, \cite{2017RAALi} argued
that the fluence follows a simple power-law distribution, i.e.
$N(F) \propto F^{-1.1 \pm 0.2}$. Our current results differ
markedly from that of \cite{2017RAALi}. Especially, we now have
three segments in the distribution, but not a single power-law
component. The main reason that leads to the difference is the
sample size. \cite{2017RAALi}'s sample only includes 16 bursts
detected at an early stage of the FRB field. Those events are all
relatively strong bursts, which may actually correspond to the
bright segment of our current distribution. It is interesting to
note that our index of $\alpha_3 = 2.06 \pm 0.90 $ is roughly
consistent with \cite{2017RAALi}'s  value of $-1.1 \pm 0.2$ in the
error range. We also notice that the error range of our $\alpha_3$
is still very large. It reflects the fact that the number of high
fluence FRBs are still too small, which can also be clearly seen
in the left panel of \autoref{fig:best-fitted}. More high fluence
FRBs are necessary to better constrain this index in the future.

The parameters derived from the fluence distribution can provide
useful insights into the nature of non-repeating FRBs. The steeply
increasing low-fluence segment, characterized by the index
$\alpha_1$, may reflect the sensitivity limit of the CHIME
telescope, which leads to the complete absence of very distant
events as well as some nearer but intrinsically weaker bursts.
Note that for those bursts whose fluence is only slightly above
the sensitivity limit, the signal could also be weakened by
scattering and dispersion effects of electrons in the
intergalactic medium
(\text{IGM})~\citep{Macquart2018,Shannon2018}.
The Eddington-type biases may also play a role
here, which refer to the overestimation of faint source counts
near the sensitivity limit when upward noise fluctuations push
marginal sources above threshold more often than downward ones
\citep{Eddington1913,Crawford1970}. Such biases are common in
flux-limited transient searches and galaxy surveys. Consequently,
the steep rise in this segment could be a combination of various
factors involving sensitivity and Eddington-type biases near the
detection threshold, and scattering/dispersion smearing of
marginal events. We therefore caution against interpreting the
low-fluence slope as intrinsic (see also \citealt{2021CHIME/FRB}).

In contrast, in the high-fluence region, we derive the power-law
index as $\alpha_3 = 2.06 \pm 0.90$, which means a quick decline
of high fluence bursts. For these strong bursts, it is unlikely
that they could be omitted by the telescope. In other words, the
sensitivity limit would not affect the observations in this
region. However, extreme energetic events are usually rare, so we
only have very limited number of events in this segment. It leads
to a large fluctuation in the FRB count in each fluence bin, and
consequently a large error in $\alpha_3$. These rare events might
be accompanied by unique electromagnetic counterparts or
high-energy photons/particles, highlighting their potential as
probes of extreme physical conditions near magnetars
~\citep{2020Bochenek,2020MargalitMN}. In the future, a
significantly expanded sample will help determine $\alpha_3$
accurately, which is crucial for revealing the nature of
non-repeating FRBs.

\begin{table}
\centering \caption{  Average values of the
best-fit parameters for the filtered sample by using a
three-segment power-law function. }
\begin{tabular}{lcc}
\hline \textbf{Parameter} & \textbf{Value} & \textbf{Uncertainty} \\
\hline
$a$ & {69.09} & {$\pm 21.61$} \\
$F_{\rm 1}$ (\rm Jy ms) & {1.55} & {$\pm 0.40$} \\
$F_{\rm 2}$ (\rm Jy ms) & {11.23} & {$\pm 3.89$} \\
$\alpha_1$ & {-3.76} & {$\pm 1.61$} \\
$\alpha_2$ & {0.20} & {$\pm 0.68$} \\
$\alpha_3$ & {2.06} & {$\pm 0.90$} \\
$\omega$ & {1.32} & {$\pm 1.09$} \\
\hline
\end{tabular}
\label{tab:best_fit_params}
\end{table}

In the mid-fluence range, the distribution is interestingly
relatively flat, with $\alpha_2 = 0.20 \pm 0.68$. This segment,
which is unlikely too seriously affected by the sensitivity limit
as well as by fluctuations, may reflect the intrinsic population
features of non-repeating FRBs. We notice that the width of this
segment is quite narrow. It mainly ranges in 1.5 -- 11.2 Jy ms
(see the parameters $F_{\mathrm 1}$ and $F_{\mathrm  2}$ in
Table~\ref{tab:best_fit_params}), which somewhat indicates that
the energetics of non-repeating FRBs is clustered. As a result, a
unified energy release mechanism is hinted, which is consistent
with the seemingly one-off characteristics of these non-repeating
events. It strongly points to the hypothesis that non-repeating
FRBs are produced by catastrophic activities
\citep{2020MargalitMN,Margalit2020APjL}.

\section{\text{DM}, redshift and energy distributions}
\label{sec:method}

\begin{figure*}
\centering
\includegraphics[width=0.49\textwidth]{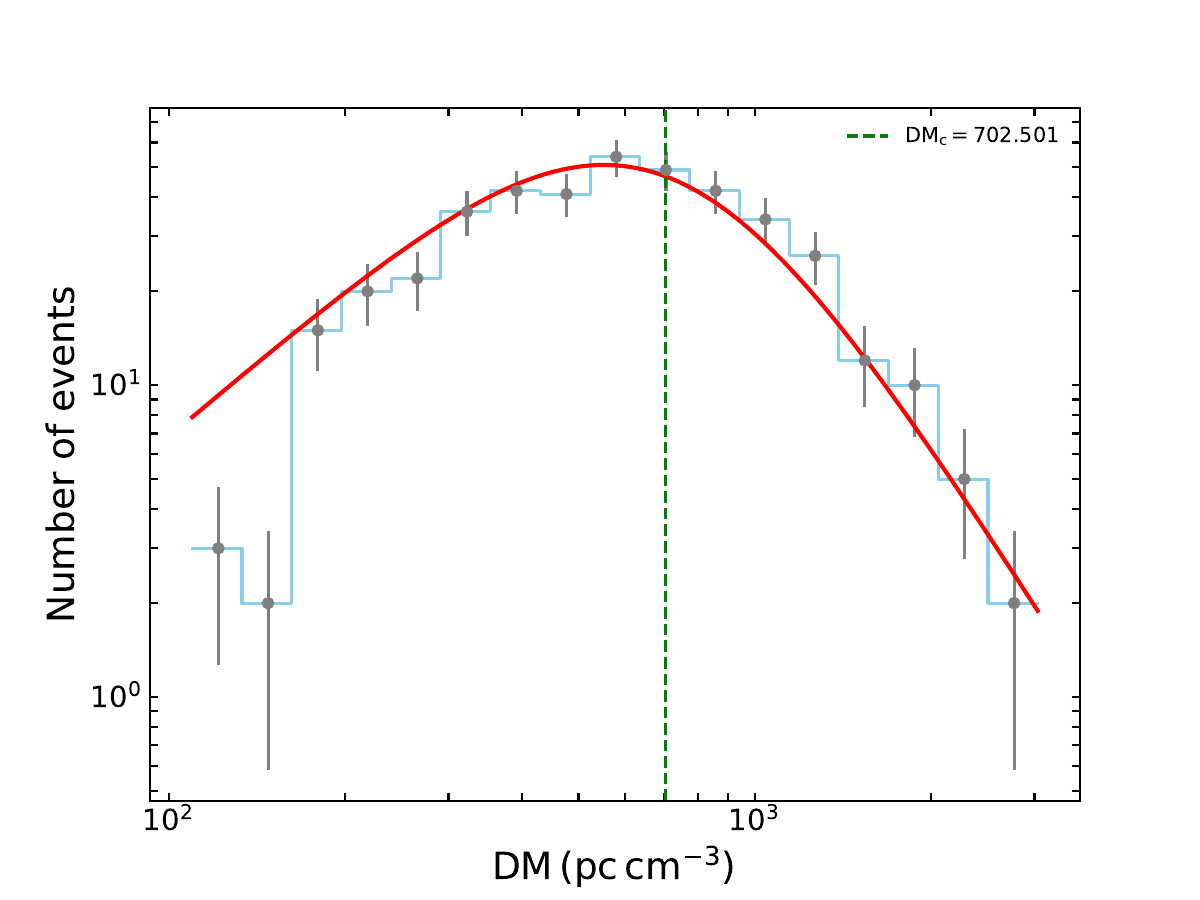}
\includegraphics[width=0.49\textwidth]{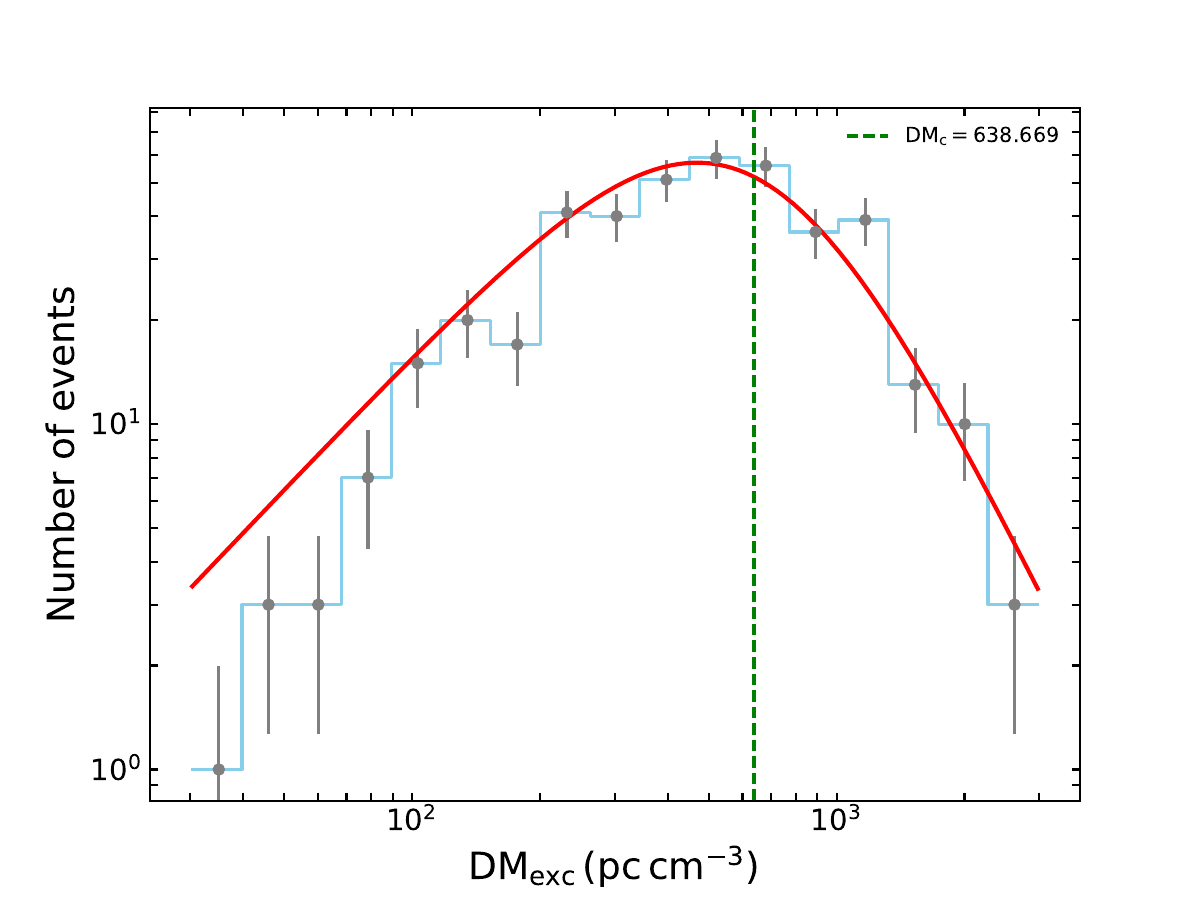}
\caption{The distribution of dispersion measure for non-repeating
FRBs in the filtered sample. The left panel shows the total
$\mathrm{DM}$ distribution, while the right panel displays the
distribution of $\mathrm{DM}_{\text{exc}}$, which is obtained by
subtracting the Galactic contribution from $\mathrm{DM}$. Note
that the Galactic contribution is calculated by referencing to the
NE2001 and YT20 models. Error bars indicate 1$\sigma$ statistical
uncertainty in each bin. A smoothly broken power-law function (the
solid curve) is used to fit the histogram in each panel.
The vertical dashed line marks the
 $\mathrm{DM}_{\mathrm{c}}$
parameter of this model, which roughly corresponds to the break
point of the best fit curve. }
 \label{fig:best-fitted DM}
\end{figure*}

\begin{table*}
\centering
\caption{Parameters derived by fitting the dispersion
measure distribution with a smoothly broken power-law function}
\begin{tabular}{lcccc}
\hline
\textbf{Parameter} & \textbf{Total \text{DM} } & \textbf{Uncertainty} &
\textbf{$\mathrm{DM}_{\rm exc}$} & \textbf{Uncertainty} \\
\hline
Normalization factor \(b\) & {162} & {$\pm 14$} & {177} & {$\pm$12} \\
$ \mathrm{DM}_{\text{c}}$ (\(\text{pc}~\text{cm}^{-3}\)) & {703} & {$\pm 14$} & {639} & {$\pm 16$} \\
$\beta_1$ & {-1.62} & {$\pm 0.04$} & {-1.30} & {$\pm 0.03$} \\
$\beta_2$ & {3.01} & {$\pm 0.13$} & {2.54} & {$\pm 0.11$} \\
$\delta$ & {1.56} & {$\pm 0.04$} & {0.57} & {$\pm 0.04$} \\
\hline
\end{tabular}
\label{tab:fitted_parameters}
\end{table*}

The observed DM of an FRB is actually a sum of contributions from
several components, including the host galaxy and source region
~\citep{Thornton2013,Deng2014,Prochaska2019}, i.e.,
\begin{equation}
\mathrm{DM} = \mathrm{DM}_{\text{MW}} + \mathrm{DM}_{\text{halo}} + \mathrm{DM}_{\mathrm{IGM}} +
\frac{\mathrm{DM}_{\mathrm{host}} + \mathrm{DM}_{\mathrm{src}}}{1+z},
\end{equation}
where $\mathrm{DM}_{\text{MW}}$ and $\mathrm{DM}_{\text{halo}}$ represent the
contributions from the Milky Way disk and halo, respectively. $
\mathrm{DM}_{\text{IGM}}$ denotes the contribution from the intergalactic
medium, and $\mathrm{DM}_{\rm {host}} + \mathrm{DM}_{\rm {src}}$ accounts for the
contributions from the host galaxy and the source region, with the
redshift factor $1+z$ applied to account for cosmological effects.
The extragalactic dispersion measure, $\mathrm{DM}_{\rm {exc}}$, is defined
as
\begin{equation}
\mathrm{DM}_{\mathrm{exc}} = \mathrm{DM}_{\mathrm{IGM}} + \frac{\mathrm{DM}_{\mathrm{host}} +
\mathrm{DM}_{\mathrm{src}}}{1+z}.
\label{eq:quadratic_formula5}
\end{equation}
$ \mathrm{DM}_{\rm {MW}}$ can be calculated by using two widely adopted
Galactic electron density models, NE2001 and YMW16. The NE2001
model ~\citep{Cordes2002,Cordes2003} is based on pulsar
observations, which describes the electron density distribution
within the Milky Way, incorporating the contributions from the
Galactic disk, spiral arms and local cloud structures to provide a
precise estimate of the dispersion measure along specific lines of
sight. YMW16 \citep{Yao2017} is a newly developed model which
utilizes more recent pulsar data and improved assumptions about
the Galactic structure, offering an enhanced accuracy for certain
directions. The stability and consistency of the NE2001 model
across a broad range of lines of sight make it more appropriate
for the precise estimation of $\mathrm{DM}_{\text{MW}}$ in this study
~\citep{Cordes2016}. So, we use the NE2001 model to estimate
$\mathrm{DM}_{\text{MW}}$ in our calculations. Another important component,
$\mathrm{DM}_{\text{halo}}$, the contribution from the Galactic halo, is
calculated based on the YT20 model \citep{Yamasaki2020}.

In our calculations, we use the open-source Python tool
PyGE\text{DM}~\footnote{https://pygedm.readthedocs.io/en/latest/contents.html}
to calculate the Galactic \text{DM} contribution. This tool
integrates the NE2001, YMW16, and YT20 models~\citep{Price2021},
allowing for efficient computation of both the Galactic dispersion
measure ($\mathrm{DM}_{\text{MW}}$) and the halo contribution
($\mathrm{DM}_{\text{halo}}$). To estimate
$\mathrm{DM}_{\text{IGM}}$, we adopt the
redshift-$\mathrm{DM}_{\text{IGM}}$ relation of
$\mathrm{DM}_{\text{IGM}} \sim 855 z ~\rm{pc}~\rm{cm}^{-3}$
\citep{Zhang2018,Macquart2020}, which is based on the latest
cosmological parameters in the flat $\Lambda$CDM model.
Specifically, the Hubble constant is taken as $H_0 = 67.8$ km
s$^{-1}$ Mpc$^{-1}$, the matter density is $\Omega_m = 0.308$, and
the dark energy density is $\Omega_\Lambda = 0.692$
\citep{2020A&ACollaboration}. The IGM is assumed to contain
approximately $f_{\text{IGM}} \sim 0.83$ of the baryons, and both
hydrogen and helium are assumed to be fully ionized up to a
redshift of $z \sim 3$. The expression is particularly useful for
constraining the redshifts of FRBs and estimating the contribution
of the IGM to the baryon density of the universe.

For simplicity, the contribution from the host galaxy and the
local source region is assumed to be a constant as $ \mathrm{DM}_{\rm
{host}} +  \mathrm{DM}_{\rm {source}} \sim
100~\rm{pc}~\rm{cm}^{-3}$~\citep{Zhang2023}. In this way, the
redshift $z$ of each FRB is estimated. It is notable that the
\citet{Zhang2018}'s $\mathrm{DM}$ - $z$ calibration itself carries an
average $\sim6\%$ statistical uncertainty. Therefore, we perturb
each $\mathrm{DM}_{\rm exc}$ by $\sim6\%$, solve Eq.~(3) for the
corresponding $z_{\rm high}$ and $z_{\rm low}$, and propagate this
range through $D_{\rm L}(z)$ to obtain the quoted errors on both
$z$ and $E_{\rm {iso}}$. In this process, we adopt the maximum
deviation from the central value as the error estimate for each
parameter. When visualizing the distributions of $z$ and $E_{\rm
iso}$, we use the central values. This approach provides a
conservative uncertainty estimation, while keeping the
presentation of the main distributions clear and concise. The
relationship between $ \mathrm{DM}_{\rm {IGM}}$ and $z$ allows the observed
dispersion measure to be connected to the redshift of the FRB,
providing a useful method for estimating $ E_{\text{iso}}$. After
obtaining an estimation of $z$, the isotropic equivalent energy
$E_{\rm iso}$ of the burst can be calculated as
\begin{equation}
E_{\rm iso} \sim \frac{4\pi D_\text{L}^2 F \nu_c}{1+z},
\label{eq:quadratic_formula6}
\end{equation}
where $\nu_\text{c}$ is the central frequency of the observational
band, and $D_\text{L}$ is the luminosity distance, which is
derived from the redshift $z$ based on the flat $\Lambda$CDM
cosmology model. It should be noted that the assumption of
isotropic radiation may lead to an overestimation of the true
energies, owing to the coherent nature of FRB emission and
potential beaming effects \citep{Katz2024}.
We nevertheless use
isotropic equivalent energies because FRB beaming angles are
poorly constrained for most sources, and introducing de-beaming
would add model dependent assumptions. Using the same isotropic
conversion for the whole sample largely preserves the
distributional shape (an unknown beaming factor predominantly
broadens the energies in log space), so our model selection mainly
reflects the data rather than uncertain beaming corrections. A
fully beaming-corrected analysis will be valuable when better
constraints on FRB beaming become available in the future.

Note that due to largely uncertain $\mathrm{DM}_\text{host}$
contribution to the total DM, there is a
dispersion in the Macquart relation, which introduces an
approximate 6\% uncertainty in DM estimates
~\citep{McQuinn2014ApJ,Zhang2018,2022James}. This uncertainty is
propagated to redshift ($z$) and isotropic energy ($E_{\rm iso}$),
which could be estimated by using a Monte Carlo-like approach: for
each $ \mathrm{DM}_{\rm exc}$, we first solve the quadratic approximation
of the Macquart relation for the central $z$ value. Then the
upper/lower bounds can be derived by varying $\mathrm{DM}$ by $\pm$6\% and
selecting positive roots (detailed in our computational procedure
below). The resulting $z$ error is further propagated to $E_{\rm
iso}$ via the luminosity distance formula,
ensuring useful error estimates for the isotropic energy.

Following the procedure described above, we have calculated
$\mathrm{DM}_{\text{exc}}$, $z$, and $E_{\text{iso}}$ for each FRB. The
original observational data and the calculated parameters are
listed in the Appendix. ~\autoref{fig:best-fitted DM} shows the
distribution of the dispersion measure of these non-repeating
FRBs, where the left panel corresponds to the total $\mathrm{DM}$, and the
right panel corresponds to $\mathrm{DM}_{\text{exc}}$.
 \cite{2021CHIME/FRB} used a lognormal
distribution to fit the DM distribution of FRBs. However, we
notice that the DM profiles in ~\autoref{fig:best-fitted DM} are
unsymmetrical. Especially, the profile has a high-DM tail
morphology that does not become flat in the large DM regime.
Theoretically, a power-law like function is supported due to at
least two reasons: (i) the comoving volume increases approximately
as a power-law with distance \citep{Macquart2018}; (ii) the
contributions of the IGM and host/local environments to DM
increase and broaden with redshift in a quasi power-law fashion
\citep{Macquart2020}. Furthermore, considering that the survey
completeness introduces a pronounced turnover near a
characteristic DM \citep{Shannon2018}, we finally adopt a smoothly
broken power-law form rather than a single lognormal function to
fit the DM distribution in this study. It captures both the rising
branch and the high-DM tail while yielding a well-defined
characteristic break value for DM. The fitting function takes the
form of
\begin{equation}
N(\mathrm{DM}) = b \left[ \left(\frac{\mathrm{DM}}{\mathrm{{DM}}_{\text{c}}}\right)^{\beta_1
\delta} + \left(\frac{\mathrm{DM}}{\mathrm{{DM}}_{\text{{c}}}}\right)^{\beta_2 \delta}
\right]^{-\frac{1}{\delta}},
\end{equation}
where $b$ is the normalization factor, $
{\mathrm{{DM}}_{\text{{c}}}}$
represents the characteristic \text{DM} value
that features the break point. $\beta_1$ and $\beta_2$ are the
power-law indices before and after the break, respectively, and
$\delta$ determines the smoothness of the transition between the
two regimes. The Poisson fluctuation ($\sigma =
\sqrt{\mathrm{N}_\text{i}}$) is used as the weight for each bin
during the fitting process. The covariance matrix of the
parameters are calculated, allowing us to derive the confidence
intervals for the fitted parameters. The fitting results are shown
in~\autoref{fig:best-fitted DM}. The derived parameters are
summarized in ~\autoref{tab:fitted_parameters}.
%% A smoothly broken power-law function is adopted to fit the distributions.

The vertical dashed line in
\autoref{fig:best-fitted DM} marks the $\rm {DM}_{c}$ value of the
best-fit continuous model, i.e. the break point. Because the
histogram is constructed with logarithmic binning and each bin is
subject to Poisson counting noise (scale as $\sqrt{\mathrm{N}}$),
the fitted curve does not perfectly coincide with the discrete
histogram. Especially, the peak of the fitted curve does not
necessarily coincide with the highest bin and can be offset from
it by fluctuations of order $10\%$, as expected for our sample
size \citep{Gehrels1986}.

For the total \text{DM} distribution, the power-law indices before
and after the break point are $\beta_1 =
-1.62\pm 0.04$ and  $\beta_2 = 3.01\pm 0.13$,  respectively. We
notice that the variation of the index is quite significant, i.e.
$\Delta \beta \sim 4.57$. In the case of
$\text{DM}_{\rm exc}$ distribution, we have slightly flatter
slopes in both segments, i.e. $\beta_1 =
-1.30\pm 0.03$ and $\beta_2 = 2.54 \pm 0.11$, with a smaller
variation of $\Delta \beta \sim 3.84 $. Since
$\mathrm{DM}_{\rm exc}$ is more likely connected with the intrinsic
distance of the FRB source than the total \text{DM}, such a
distribution may reflect the event rate evolution of non-repeating
FRBs in the local Universe.

\begin{figure*}
\centering
\includegraphics[width=0.9\textwidth]{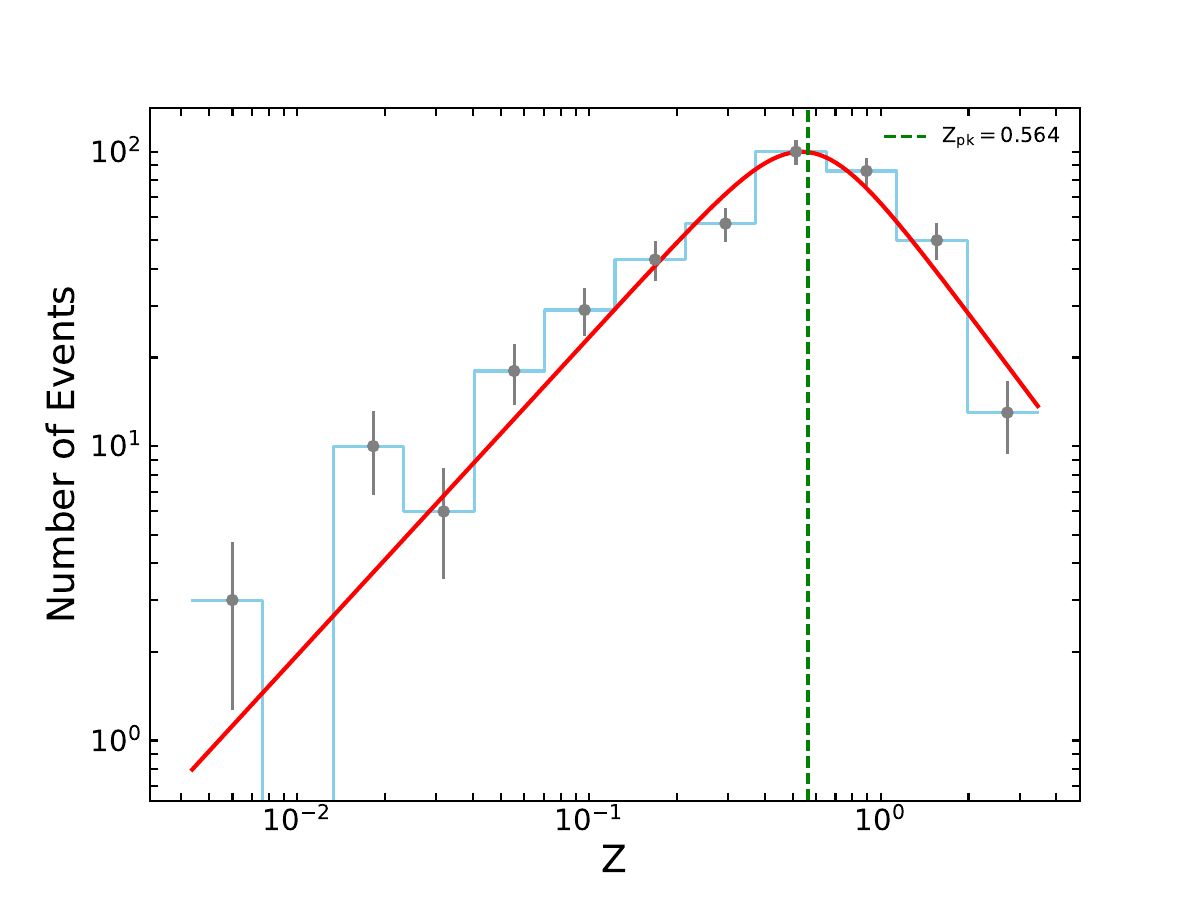}
\caption{Redshift distribution of non-repeating FRBs
in the filtered sample. Error bars indicate
1$\sigma$ statistical  uncertainty in each bin. The solid curve
shows the best-fit result by using a smoothly broken power-law
function. The vertical dashed line marks the peak of the fit
curve, which corresponds to a redshift of ~ $z = 0.564$.}
\label{fig:FRB_redshift_distribution}
\end{figure*}

The redshift distribution of non-repeating FRBs is shown in
~\autoref{fig:FRB_redshift_distribution}, along with a best-fit
curve by using a smoothly broken power-law function. We see that
the distribution peaks at a redshift of
$z_{\rm pk} = 0.564$. In the low redshift regime, the power-law index is $-1.08
\pm 0.21$, which indicates that the event rate increases as
redshift increases. This behavior is similar to that of star
formation rate to some extent~\citep{Madau1998,
Hopkins2006,2014ARA&A,Bannister2019}, supporting the idea that
non-repeating FRBs may be connected to catastrophic events related
to the death of massive
stars~\citep{Yamasaki2018,Cordes2019ARAA,Hashimoto2020a}.
However, note that the low-z increase might
also be affected by observational bias and even the uncertainties
in the Macquart relation ~\citep{2022James}. At low redshift, our
Galaxy and the host galaxy contribute a significantly larger
portion in the total DM, which may lead to a larger error in the
estimated redshift. In the future, a much larger and more
complete sample would help clarify this issue.

In contrast, in the high-redshift segment ($z > z_{\text{pk}}$),
the power-law index is $1.33 \pm 0.60$, indicating a steep decline
in the number of observed sources. Interestingly, similar results
are also reported by other
groups~\citep{Zhang2020,Macquart2020,Hashimoto2020,Zhang2021}.
While such a decline might be partially caused by observational
bias that distant FRBs are usually much difficult to be detected,
it could also reflect some intrinsic features of the FRB engines.
For example, it might be driven by the frequent formation of
massive galaxies and other physical processes occurring in the
earlier cosmic epochs~\citep{Behroozi2013,Peng2025}.

\begin{figure*}
\centering
\includegraphics[width=0.9\textwidth]{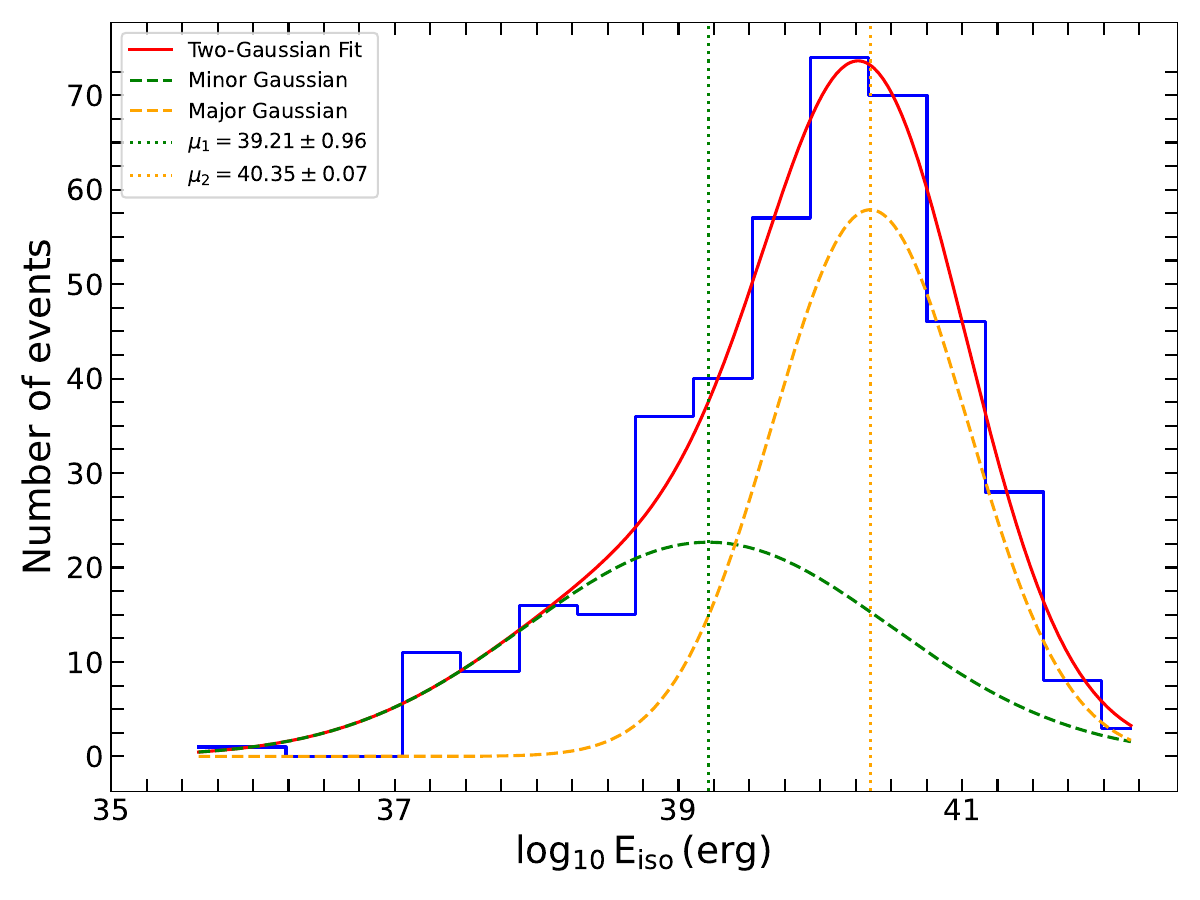}
\caption{ The distribution of isotropic energy release of
non-repeating FRBs in the filtered sample. The solid curve shows the best-fit result by
engaging two Gaussian components, with each component shown by the
dashed curve separately. The green dashed curve corresponds to the
minor Gaussian component centered at $\mu_1 = 39.21 \pm 0.96$, with
a dispersion of $\sigma_1 = 1.29 \pm 0.37$. The orange dashed curve
corresponds to the major Gaussian component centered at $\mu_2 =
40.35 \pm 0.07$, with a dispersion of $\sigma_2 = 0.69 \pm 0.13 $ . }
\label{fig:Eiso_distribution}
\end{figure*}

The energy distribution of the filtered sample is shown in
~\autoref{fig:Eiso_distribution}. We see that the histogram is
asymmetric, which obviously could not be matched by a single
Gaussian component. To quantitatively determine the best
description of the data, we tested several empirical models,
including a simple power-law, a broken power-law, a
single-Gaussian, and a two-Gaussian mixture. Model selection was
performed using the Akaike Information Criterion (AIC;
\citet{Akaike1974}) and the Bayesian Information Criterion (BIC;
\citet{Schwarz1978}), which penalize unnecessary model complexity
while rewarding better fits. As summarized in
Table~\ref{tab:model_comp_energy}, the two-Gaussian model achieves
the lowest AIC and BIC values, outperforming the single-Gaussian
and power-law alternatives by $\Delta$AIC $\approx$ 8 -- 60 and
$\Delta$BIC $\approx$ 6 -- 57, respectively. We therefore adopt
the two-component interpretation for $E_{\rm iso}$ in the
following analysis.
We note that the low-fluence rise shown in Figure 1, potentially
influenced by selection effects and biases as discussed in Section
3, may propagate to the energy distribution via fluence in
Equation (4). Nevertheless, the narrow clustering of the major
component and the built-in penalty of AIC/BIC against extra
parameters suggest that overfitting is unlikely to drive the
bimodal structure. Future larger samples could help further test
it.

\begin{table}
\centering
\small
\caption{Model comparison for the isotropic energy distribution of non-repeating FRBs.
Lower AIC/BIC indicate a statistically preferred model; the two-Gaussian model is favoured.}
\label{tab:model_comp_energy}
\begin{tabular}{lcccc}
\hline
\textbf{Model} &
\shortstack{\textbf{Parameter}\\\textbf{number}} &
\textbf{AIC} & \textbf{BIC} &
\shortstack{\textbf{$\Delta$AIC} / \\ \textbf{$\Delta$BIC}} \\
\hline
Double Gaussian   & 6 & \textbf{46.835} & \textbf{51.083} & --- / --- \\
Broken power-law  & 4 & 54.980 & 57.013 & +8.15 / +5.93 \\
Single Gaussian   & 5 & 62.060 & 64.984 & +15.23 / +13.90 \\
Simple power-law  & 2 & 107.111 & 108.527 & +60.28 / +57.44 \\
\hline
\end{tabular}
\end{table}

The adopted fitting function takes the form of
\begin{equation}
f(x) = \text{c}_1 \exp\left(-\frac{(x - \mu_1)^2}{2\sigma_1^2}\right) +
\text{c}_2 \exp\left(-\frac{(x - \mu_2)^2}{2\sigma_2^2}\right).
\end{equation}
Here \(c_1\) and \(c_2\) are two coefficients. \(\mu_1\) and
\(\mu_2\) correspond to the central energies of the two Gaussian
components, while \(\sigma_1\) and \(\sigma_2\) are the standard
deviations. The best-fit parameters of the two components are
$\mu_1 = 39.21 \pm 0.96$, $\sigma_1 = 1.29 \pm
0.37$ and $\mu_2 = 40.35 \pm 0.07$, $\sigma_2 = 0.69 \pm 0.13 $.
In other words, the major component has a characteristic energy of
$\sim 2.3 \times 10^{40}$ erg, with a narrow
dispersion range of $\sim 4.6 \times 10^{39}$
--- $1.1 \times 10^{41}$ erg. The minor component has a slightly
smaller characteristic energy of $1.6\times
10^{39}$ erg, but with a significantly wider dispersion range of
$\sim 8.3 \times 10^{37}$ --- $3.2 \times
10^{40}$ erg. The best-fit parameters
derived from the full sample are consistent with those from the
filtered sample within $1\sigma$ uncertainty (e.g.,
$\Delta\mu_{1}\approx0.12\sigma$,
$\Delta\mu_{2}\approx0.85\sigma$, see Appendix
~\autoref{fig:Eiso_distribution1}).

From \autoref{fig:Eiso_distribution}, we can see that the number
of major component FRBs is much larger than that of the minor
component. They are essentially the predominant events. More
interestingly, they are mainly in a narrow energy range of
$4.6 \times 10^{39} {\rm erg} <  E_{\rm iso} < 1.1
\times 10^{41} {\rm erg} $, strongly pointing to a uniform energy
reservoir for them. For example, they could be giant flares from
young magnetars~\citep{Metzger2019}. But if they are confirmed as intrinsically
one-off events, then it would be more likely that they may be
triggered by some catastrophic processes such as neutron star
mergers. Anyway, the narrow distribution of $E_{\text{iso}}$
indicates that these non-repeating FRBs could act as standard
candles and may be utilized in cosmology studies. Due to their
brightness, they are detectable at large distances, making them
valuable tools for studying the large-scale structure of the
universe, the intergalactic medium, and cosmic baryon distribution
\citep{Zhang2018, Macquart2020}.

In contrast, the broader energy range of the minor Gaussian
component indicates that these lower-energy population may have a
different origin. For example, they may come from repeating FRB
sources but have only been detected for one time. It is
interesting to note that in a comprehensive analysis of repeating
FRBs, \cite{Hu2023} have shown that the energies of the bursts
from those active repeating sources generally span over two
magnitudes. The lower-energy population here are quite similar to
those repeating FRBs in this aspect.

\section{Comparison with Previous Studies}
\label{sect:Comparison}

Our results on the fluence and energy
distributions of non-repeating FRBs generally align with several
previous studies. For the fluence distribution, our three-segment
power-law result (with $\alpha_1 \approx -3.76 \pm 1.61$,
$\alpha_2 \approx 0.20 \pm 0.68$, $\alpha_3 \approx 2.06 \pm
0.90$) can be regarded as an extension of earlier single power-law
fit. For instance, \cite{2017RAALi} reported a power-law index of
$-1.1 \pm$ 0.2 by using a single power-law component to fit a
small sample of 16 non-repeating FRBs. We notice that our
$\alpha_3$ value is roughly consistent with their index in the
uncertainty range, which strongly indicates that their early
sample was mainly comprised of bright events, while our sample is
much larger and includes low- and mid-fluence burst due to
improved telescope sensitivity. More recently, \cite{James2019}
identified a steepening in the fluence distribution at a threshold
of $\sim$ 5 -- 40 Jy ms. Our transition points ($F_{\rm 1} \approx
1.56$ Jy ms, $F_{\rm 2} \approx 11.30$ Jy ms) are consistent with
their results,showing that the multiple-segment structure is an
intrinsic feature of the distribution rather than due to
observational biases.
However, the steep rise at the
low-fluence end is likely dominated by selection effects near the
detection threshold (Eddington bias and dispersion/scattering
smearing), and thus should be interpreted with caution as it may
not fully represent an intrinsic population property
(see also \citealt{2021CHIME/FRB}).

The energy distribution has been explored in
several studies. Early analyses of mixed FRB populations
(repeating and non-repeating sources) with limited samples
suggested a power-law energy distribution. For example,
~\cite{Lu2019L} reported an index of $\sim -1.8$. In contrast,
~\cite{Luo2018} measured the normalized luminosity function of
FRBs using a Bayesian approach and found slopes between $\sim
-1.8$ and $-1.2$. Using the CHIME/FRB catalogue with calibrated
selection effects, ~\cite{Shin2023ApJ} inferred a Schechter energy
function for the FRB population. \cite{Bhattacharyya2023MNRAS}
analysed 254 non-repeating, low-DM events from CHIME/FRB Catalog~1
and modelled the isotropic-equivalent energy distribution with a
modified Schechter form, reporting a low-energy excess relative to
the high-energy tail. In addition, ~\cite{Cui2022} identified a
Gaussian-like feature in the luminosity distribution of the CHIME
sample. For the active repeater FRB~20121102A, ~\cite{Li2021Natur}
reported a bimodal energy distribution. Taken together, these
studies point toward multi-component behaviour. Using our
substantially enlarged CHIME non-repeater sample, we find that the
isotropic-equivalent energy distribution is best described by a
two-Gaussian mixture, further indicating that single-component
models are inadequate for FRB energetics.

We note that the apparent difference in sample size of our data
set as compared with others arises from our use of
\texttt{fitburst} SNR instead of the \texttt{bonsai} trigger SNR
for event selection (see ~Section~\ref{sect:Obs}). This choice
allows inclusion of morphologically broader or narrower-band
bursts whose true amplitudes are underestimated in \texttt{bonsai}
SNR.

\section{Conclusions and Discussion}
\label{sec:conclusions}

In this study, we present a comprehensive analysis on 415
non-repeating FRBs observed by CHIME,
focusing on the key parameters such as $F$, $\mathrm{DM}$,
$\mathrm{DM}_{\text{exc}}$, $z$, and $E_{\text{iso}}$. It is found that the
fluence distribution can be modeled by a three-segment broken
power-law function. The power-law index in each segment is
$\alpha_1 = -3.76 \pm 1.61$, $\alpha_2 = 0.20
\pm 0.68$, and $\alpha_3 = 2.06 \pm 0.90$ , respectively. The
mid-fluence segment, which spans only in a narrow fluence range
(1.5 -- 11.2 Jy ms), is notably quite flat.
The distributions of both total $\mathrm{DM}$ and $\mathrm{DM}_{\text{exc}}$ can be
well fitted by two-segment smoothly broken power-law functions.
For the $\mathrm{DM}_{\rm exc}$ distribution, the two power-law indices are
$\beta_1 = -1.30 \pm 0.03$ and $\beta_2 = 2.54
\pm 0.11$, respectively, with a variation of $\Delta \beta \sim
2.97$. The IGM contribution to the dispersion measure
($\mathrm{DM}_{\text{IGM}}$) is used to estimate the redshift of each FRB
based on the Macquart relation, which naturally leads to an
estimation of the isotropic energy release. It is found that the
redshift distribution peaks at $z_{\rm pk} =
0.564$. In the low redshift regime, the power-law index is
$-1.08 \pm 0.21$, while it is
$1.33 \pm 0.60$ in the high-redshift segment.

Interestingly, $E_{\text{iso}}$ seems to exhibit a bimodal
distribution, which includes two Gaussian components. The higher
energy component is the dominant population, which has a typical
energy of $2.3 \times 10^{40} {\rm erg}$ and
spans mainly in a narrow range of $4.6 \times
10^{39} {\rm erg} < E_{\rm iso} < 1.1 \times 10^{41} {\rm erg} $.
The narrowness of this major component strongly indicates a
uniform origin for them. On the contrary, the lower energy
component, which is a minor population, has a characteristic
energy of $1.6 \times 10^{39}$ erg and spans
over two magnitudes, i.e. $8.3 \times 10^{37}
{\rm erg} < E_{\rm iso} < 3.2 \times 10^{40} {\rm erg} $. These
events may have a multiple origin and could even come from
repeating FRB sources.

Our statistical fits to the CHIME non-repeater population indicate
that a two-component Gaussian model best describes the
isotropic-equivalent energy distribution. This bimodal structure
could reflect either (i) two distinct progenitor channels, such as
a combination of magnetar-driven bursts and magnetospheric
coherent radiation
\citep{Platts2019,Metzger2019,Lyubarsky2020,2020Lu},
or (ii) a single engine with varying radiative regimes or
geometric states, e.g., state changes and beaming effects
\citep{Kumar2017,Margalit2019}. In either
case, the dominant higher-energy component, narrowly clustered at
around $2.3 \times 10^{40}$ erg, suggests a relatively uniform
energy release mechanism, constraining progenitor models to those
favoring catastrophic events with standardized energy reservoirs
~\citep{Metzger2019,2020MargalitMN,Margalit2020APjL,Katz2024}.
The broader lower-energy component implies greater diversity,
potentially indicating heterogeneous origins within non-repeating
FRBs. Applying the same isotropic conversion preserves the
relative structure of the distribution, allowing our inference to
focus on its shape rather than absolute normalization.

Our results align with recent studies using
CHIME samples that recover Schechter-like energy functions,
including a high-energy Schechter tail and low-energy excess
observed in non-repeating FRBs \citep{Bhattacharyya2023MNRAS} and
apparent non-repeaters \citep{Shin2023ApJ}. This pattern mirrors
bimodal distributions observed in active repeating FRBs, such as
FRB 20121102A \citep{Li2021Natur}, suggesting a possible physical
connection.  For example, the lower-energy component may include
contributions from repeat-capable engines observed only once due
to observational biases ~\citep{Hu2023,Yamasaki2024MNRAS}. We
treat this two-Gaussian mixture as a working model for FRB
energetics and emphasize testable predictions: if the
higher-energy component corresponds to shock-powered emission, it
may exhibit larger rotation measures or stronger scattering tails
\citep{Margalit2020APjL,Katz2024};
alternatively, if the components reflect magnetospheric vs. shock
regimes of the same engine, differences in repetition statistics,
spectral indices, and burst morphology should emerge
\citep{Metzger2019,2020Lu,Beloborodov2020,2023Zhang}.
Future work incorporating localization-enabled redshifts,
completeness corrections, and multi-wavelength data will enable
stricter tests, including comparisons of host/ISM properties,
spectral-temporal morphology, and energy versus waiting-time
correlations across the components.

Several challenges still remain in our study. First, the redshifts
are not directly measured for our sample. Instead, the parameter
is estimated from an empirical relation, which may lead to a large
error in the parameter value. The lack of precise redshift
measurements limits the accuracy of energy and distance estimates.
Future localization and host galaxy observations will be necessary
for further refining these analyses.

Second, observational biases, such as telescope sensitivity and
selection effects, require further investigation. Our analysis is
based on CHIME observations and the sample only includes 415
non-repeating FRBs. To overcome these problems, a significantly
expanded sample is necessary. Additionally, CHIME operates mainly
in 400 MHz -- 800 MHz. It is quite unclear whether non-repeating
FRBs have similar or completely different features in other
wavelength ranges. Future observations with higher sensitivities
and broader sky/frequency coverage will be essential for a more
in-depth study. It is also important to acknowledge
several well-known observational biases and limitations inherent
to the CHIME/FRB dataset and our analysis. First, the fluence
measurements typically assume each burst occurs at the beam
center, leading to systematic underestimation of true fluence and
thus the isotropic energy ~\citep{2021CHIME/FRB}. Second, strong
selection effects exist due to the instrument's detection
threshold: FRBs with low signal-to-noise ratio (S/N $\lesssim$
15), high \text{DM}, or wide pulse widths are less likely to be detected
\citep{Merryfield2023}. These factors bias the observed fluence
and \text{DM} distributions. It should be noted that, although recent
studies suggest the \text{DM}--$z$ relation has a 30$\sim$40\% intrinsic
scatter ~\citep{Zhang2018,Macquart2020,2022James,Ma2025},
we have not incorporated this into our error propagation
in the present analysis. As such, the uncertainties reported here
are lower limits, and the real uncertainty may be correspondingly
larger.

The distinction between repeating and non-repeating FRBs is also
an interesting issue. However, the sample size of repeating FRBs
is much smaller, which makes it impossible to perform a meaningful
comparison. In the future, when more and more repeating FRB
sources are discovered, it would be necessary to compare them with
non-repeating FRBs. This will help improve the methodology of FRB
classification, and clarify their different nature as well.
Exploring the connection between FRBs and other high-energy
transients, such as gamma-ray bursts, kilonovae, gravitational
wave events, could also provide new insights on FRBs and open new
opportunities for multi-messenger astrophysics.

\section*{Acknowledgements}
This study was supported by the National Natural Science
Foundation of China (Grant Nos. 12233002, 12041306, 12447179,
12273113), by the National Key R\&D Program of China (No.
2021YFA0718500),  by the Major Science and Technology Program of
Xinjiang Uygur Autonomous Region (No. 2022A03013-1), and by the
Postgraduate Research \& Practice Innovation Program of Jiangsu
Province (No. KYCX25\_0197). YFH also acknowledges the support
from the Xinjiang Tianchi Program. A.K. acknowledges the support
from the Tianchi Talents Project of Xinjiang Uygur Autonomous
Region. Jin-Jun Geng acknowledges support from the Youth
Innovation Promotion Association (2023331).

%% Revised by HYF on: 20250425

%%%%%%%%%%%%%%%%%%%%%%%%%%%%%%%%%%%%%%%%%%%%%%%%%%

\section*{Data Availability}

The fast radio burst sample used in this study is based on publicly
available data released by the CHIME/FRB Collaboration.
The catalog can be accessed at \url{https://www.chime-frb.ca/catalog}
or \url{https://blinkverse.zero2x.org/\#/availability}.
Derived parameters such as redshift and isotropic energy were calculated
from published quantities using the methods described in Section~\ref{sec:method}.

%%%%%%%%%%%%%%%%%%%% REFERENCES %%%%%%%%%%%%%%%%%%

% The best way to enter references is to use BibTeX:

\bibliographystyle{mnras}
\bibliography{references} % if your bibtex file is called example.bib

%%%%%%%%%%%%%%%%% APPENDICES %%%%%%%%%%%%%%%%%%%%%

\clearpage

\onecolumn

\appendix
\section*{Appendix}
\label{sect:Appendix}

\section{Data of non-repeating FRBs}
Here we present the key parameters of the non-repeating CHIME FRBs
used in this study. The full data table is
available on GitHub:
https://github.com/nurimangul/full-sample.git. In the online
table, the first column is the FRB name. Other columns correspond
to the Galactic longitude, Galactic latitude, central frequency,
total \text{DM}, fluence, \text{DM} contribution of the Galactic
halo, \text{DM} contribution of the Milky Way disk, \text{DM}
excess (i.e. extragalactic \text{DM}), redshift, and the isotropic
energy release, respectively. Note that the parameters directly
derived from observations are taken from the Blinkverse website,
while other parameters such as $ \mathrm{DM}^{\rm YT20}_{\rm halo}$,
$\mathrm{DM}_{\rm MW}^{\rm NE2001}$, $ \mathrm{DM}_{\rm exc}$, $z$ and $E_{\rm iso}$
are calculated by following the procedure introduced in the main
text.

\section{Filtering Code}
To avoid observational bias as far as
possible, we apply some criteria on the original 461 FRBs to draw
a filtered sample of 415 FRBs. The criteria is detailed in
Section~2. The filtering process is realized with the following
Python code:
\begin{verbatim}
import pandas as pd
data = pd.read_excel("fullsample.xlsx")
data = data.dropna(subset=['SNR', 'DM_SNR'])
max_dm_gal =data[['DM_NE2001']].max(axis=1)
mask = (data['SNR'] >= 12) &(data['DM_SNR'] >= 1.5 * max_dm_gal) & (data['Fluence'] >= 0.4)
complete_sample = data[mask]
complete_sample.to_excel("complete_sample.xlsx", index=False).
\end{verbatim}

\section{Sensitivity Tests}
\begin{figure*}
\centering
\includegraphics[width=0.9\textwidth]{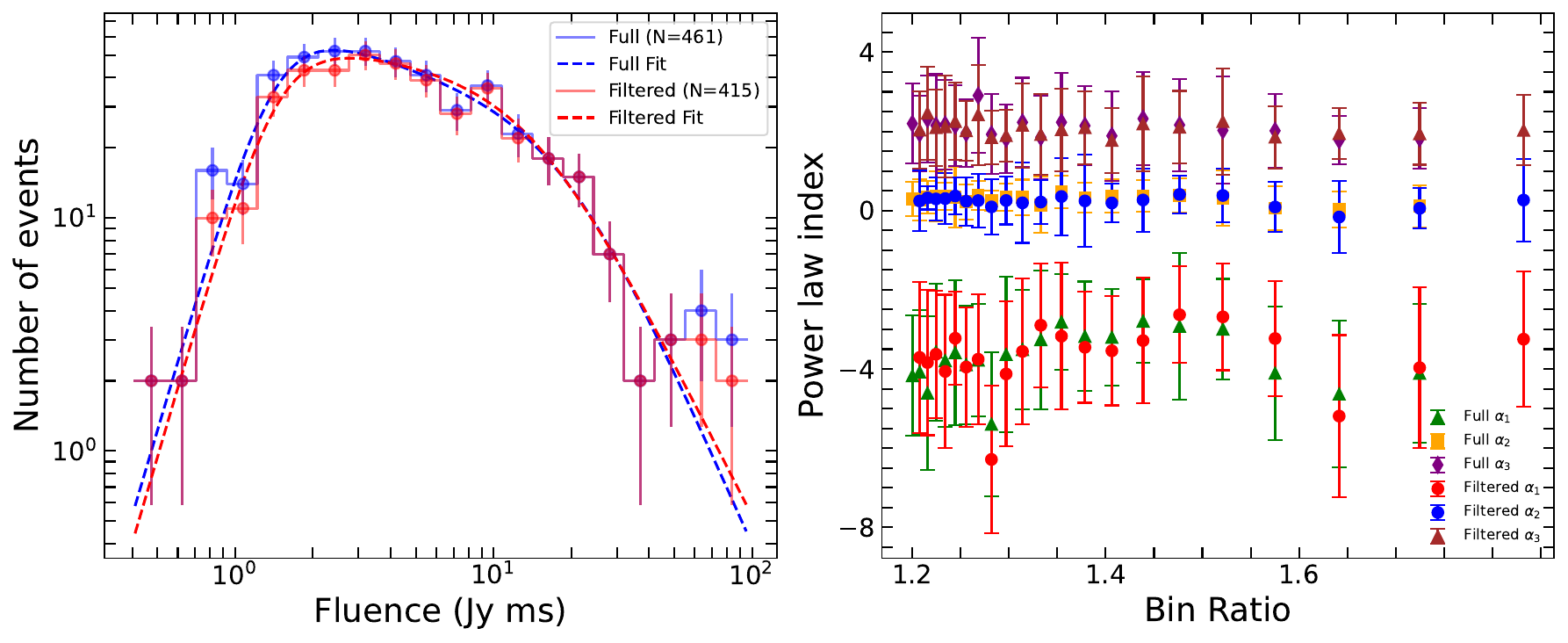}
\caption{Left panel: the fluence distribution
of non-repeating FRBs with a bin ratio of 1.3. Blue dots
correspond to original unfiltered full sample with $ \mathrm {N} = 461$
events, and red dots correspond to filtered sample with $ \mathrm {N}=415$
bursts. A best-fit curve engaging a three-segment power-law
function is also plotted for the histogram correspondingly. Right
panel: variation of $\alpha_1$,$\alpha_2$, and $\alpha_3$ with
respect to the bin ratio.}
\label{fig:dist_compare}
\end{figure*}

To see the effects of different sensitivity on
the statistics, we have compared the fluence distributions of the
original unfiltered full sample ($\mathrm{N}=461$) and the filtered sample
$(\mathrm{N}=415)$. In the left panel of ~\autoref{fig:dist_compare}, the
fluence distributions of the full sample and the filtered sample
are plotted for a bin ratio of $\sim 1.3$. The corresponding
best-fit three-segment curves are also illustrated. The best-fit
parameters for the full sample are derived as $a = 64.9 \pm 22.0$,
$F_1 = 1.5 \pm 0.2$, $F_2 = 11.9 \pm 3.9$, $\alpha_1 = -3.55 \pm
1.24$, $\alpha_2 = 0.37 \pm 0.35$, $\alpha_3 = 2.34 \pm 1.19$, and
$\omega = 1.5 \pm 1.1$. They are consistent with the corresponding
parameters derived from the filtered sample in the uncertainty
ranges. In fact, the KS test gives a statistic $D$ value of 0.039
($p=0.881$), indicating no significant difference between the full
sample and the filtered sample.

In the right panel of
\autoref{fig:dist_compare}, the indices of $\alpha_1$, $\alpha_2$,
and $\alpha_3$ are plotted versus the bin ratio. Similarly, we see
that these indices do not vary too much in the bin ratio of 1.2 --
1.8 . The average values of the best-fit parameters derived from
MCMC fitting across various bin ratios for the full sample are
presented in ~\autoref{tab:full_sample}. Again, they are
consistent with those derived based on the filtered sample (see
the main text).

\begin{table*}
\centering
\caption{Average values of the
best-fit parameters for the unfiltered full sample by using a
three-segment power-law function. }
\begin{tabular}{lcc}
\hline \textbf{Parameter} & \textbf{Value} & \textbf{Uncertainty} \\
\hline $a$ & 69.24 & $\pm 20.34$ \\
       $F_{1}$ (Jy ms) & 1.46 & $\pm 0.33$ \\
       $F_{2}$ (Jy ms) & 10.82 & $\pm 3.86$ \\
       $\alpha_1$ & -3.87 & $\pm 1.61$ \\
       $\alpha_2$ & 0.26 & $\pm 0.53$ \\
       $\alpha_3$ & 2.07 & $\pm0.96$ \\
       $\omega$ & 1.44 & $\pm1.11$ \\
\hline \end{tabular}
\label{tab:full_sample}
\end{table*}

\begin{figure*}
\centering
\includegraphics[width=0.9\textwidth]{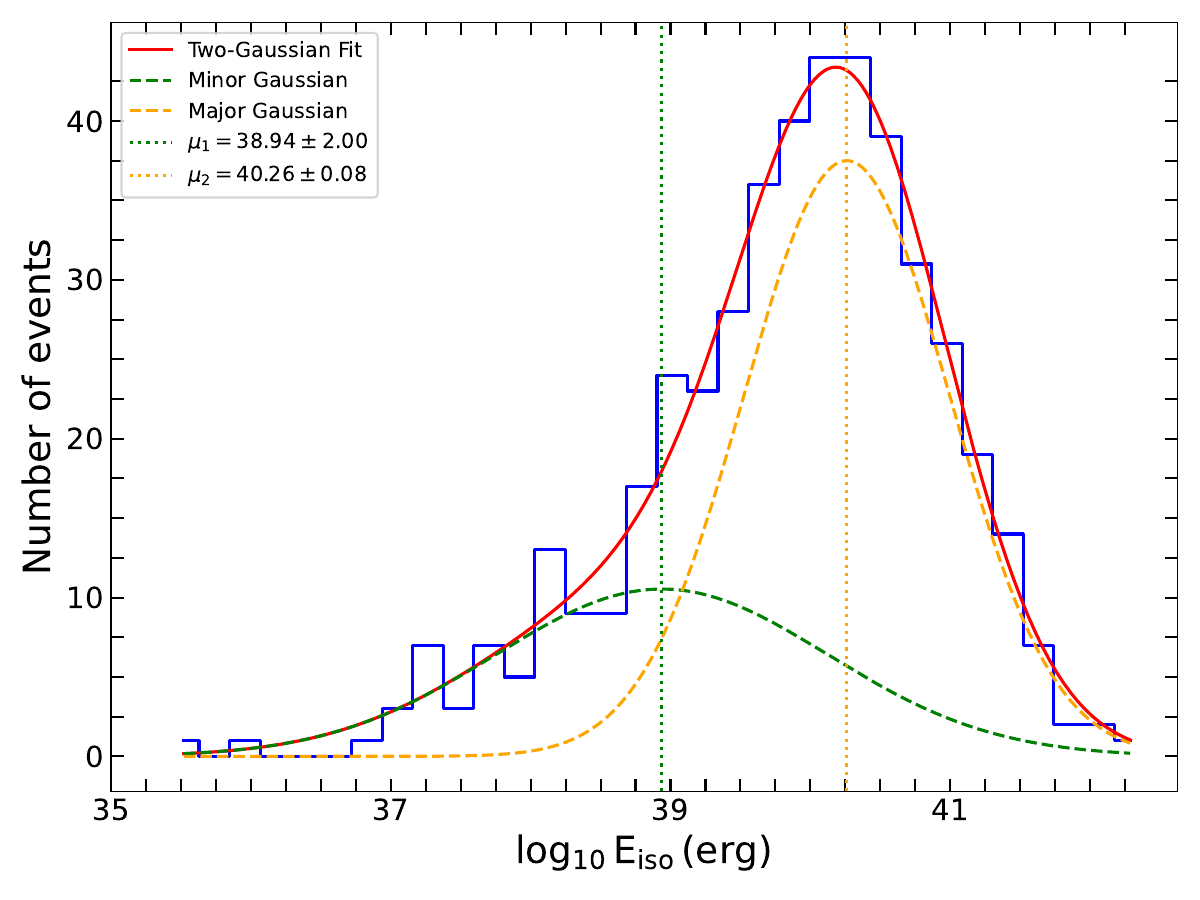}
\caption{Distribution of isotropic energies
of non-repeating FRBs in the full sample. The solid curve shows
the best-fit result by engaging two Gaussian components, with each
component shown by the dashed curve separately. The green dashed
curve corresponds to the minor Gaussian component centered at
$\mu_1 = 38.94 \pm 2.00$, with a dispersion of  $\sigma_1 = 1.19
\pm 0.86$. The orange dashed curve corresponds to the major
Gaussian component centered at $\mu_2 = 40.26 \pm 0.08$, with a
dispersion of $\sigma_2 = 0.74 \pm 0.14 $. }
\label{fig:Eiso_distribution1}
\end{figure*}

The distribution of isotropic energies of FRBs
in the full sample is plotted in
~\autoref{fig:Eiso_distribution1}. The best-fit parameters of
the two Gaussian components are $\mu_1 = 38.94 \pm 2.0$, $\sigma_1
= 1.19 \pm 0.86$ and $\mu_2 = 40.26 \pm 0.08$, $\sigma_2 = 0.74
\pm 0.14 $. In other words, the major component has a
characteristic energy of $\sim 1.8 \times 10^{40}$ erg, with a
narrow dispersion range of $\sim 3.4 \times 10^{39}$ --- $1.0
\times 10^{41}$ erg. The minor component has a slightly smaller
characteristic energy of $8.7\times 10^{38}$ erg, but with a
significantly wider dispersion range of $\sim 5.6 \times 10^{37}$
--- $1.3 \times 10^{40}$ erg. These parameters are
consistent with those derived based on the filtered sample in the
uncertainty ranges (see the main text).

\bsp    % typesetting comment
\label{lastpage}
\end{document}